\documentstyle[12pt]{article}
\topmargin - 1.5 cm % for epson - 3,5 cm. for laserjet appr. -10pt

\begin{document}

\begin{center}
{\large{\bf Hidden Symmetry of the Yang--Coulomb System}}
\end{center}
\begin{center}
{\bf L.G. Mardoyan${}^1$, A.N. Sissakian${}^2$, V.M. Ter--Antonyan${}^3$}
\end{center}
\begin{center}
{\large Joint Institute for Nuclear Research}
\end{center}
\begin{center}
141980, Dubna, Moscow Region, Russia
\end{center}

\footnotetext[1]{E-mail: mardoyan@thsun1.jinr.dubna.su}
\footnotetext[2]{E-mail: sisakian@jinr.dubna.su}
\footnotetext[3]{E-mail: terant@thsun1.jinr.dubna.su}

\vspace{0.5cm}
\begin{abstract}
{\small The bound system composed of the Yang monopole coupled
to a particle of the isospin by the $SU(2)$ and Coulomb
interaction is considered. The generalized Runge--Lenz vector
and the $SO(6)$ group of hidden symmetry are established. It is
also shown that the group of hidden symmetry make it possible
to calculate the spectrum of the system by a pure algebraic
method.}
\end{abstract}

\section {Introduction}

It is well known that gauge theory underlies modern particle physics
and gravity. One of the most interesting features of local gauge theories,
first noted by Dirac [1], is the natural occurrence in them of monopoles
as topological obstructions. As originally proposed by Yang [2], the Dirac
monopole can be generalized to the $SU(2)$ gauge group and such a
generalization (Yang monopole) can be achieved only in the five--dimensional
Euclidean space.

The simplest bound system connected to the Yang monopole is the
Yang-Coulomb System (YCS) which we define here as the system composed
of the Yang monopole and a particle of the isospin coupled to the
monopole by the $SU(2)$ and the Coulomb interaction.

It is obviously of interest to ask what happens to the known
Coulomb $SO(6)$ hidden symmetry after $SU(2)$ generalization.
In this note, we prove that $SU(2)$ generalization does not break
the $SO(6)$ Coulomb hidden symmetry but leads to $SO(6)$ acting
in a more general ${\rm I \!R}^5 \otimes S^3$ space. We use this
new symmetry for computation of the YCS energy spectrum by a pure
algebraic method.

\section {Notation and $\tau$ matrices}

We keep the following notation: $j=0,1,2,3,4$;
$\mu=1,2,3,4$; $a=1,2,3$; $x_j$ are the Cartesian coordinates of
the particle, ${\hat T}_a$ denote the $SU(2)$ gauge group generators;
${\vec A}^a=(0, A_{\mu}^a)$ is the triplet of Yang monopole's gauge
potentials; $F_{ik}^a$ is the gauge field of the Yang monopole;
${\sigma}^a$ are the Pauli matrices, and ${\tau}^a$ are the 4x4 matrices
\begin{eqnarray*}
{\tau}^1 = \frac{1}{2}\left(\begin{array}{cc}
0&i{\sigma}^1 \\
-i{\sigma}^1&0
\end{array} \right),\,\,\,
{\tau}^2 = \frac{1}{2}\left(\begin{array}{cc}
0&-i{\sigma}^3 \\
i{\sigma}^3&0
\end{array} \right),\,\,\,
{\tau}^3 = \frac{1}{2}\left(\begin{array}{cc}
i{\sigma}^2&0 \\
0&i{\sigma}^2
\end{array} \right)
\end{eqnarray*}
${\tau}^a$ matrices satify to relations
\begin{eqnarray*}
[{\tau}^a,{\tau}^b] = i{\epsilon}_{abc}{\tau}^c,\,\,\,\,
4{\tau}_{\mu \lambda}^a{\tau}_{\lambda \nu}^b =
{\delta}_{ab}{\delta}_{\mu \nu} +
2i{\epsilon}_{abc}{\tau}_{\mu \nu}^c
\end{eqnarray*}
\begin{eqnarray*}
{\epsilon}_{abc}{\tau}_{\alpha \beta}^b{\tau}_{\mu \nu}^c =
\frac{i}{2}\left({\delta}_{\alpha \mu}{\tau}_{\nu \beta}^a-
{\delta}_{\alpha \nu}{\tau}_{\mu \beta}^a+
{\delta}_{\beta \nu}{\tau}_{\mu \alpha}^a-
{\delta}_{\beta \mu}{\tau}_{\nu \alpha}^a\right)
\end{eqnarray*}

\section {Yang monopole}

Recently [3,4], the YCS has been extracted from the
eight--dimensional isotropic quantum oscillator with the help of the
special transformation ${\rm I \!R}^8 \to {\rm I \!R}^5 \otimes S^3$.
Here, we shall use the formula
\begin{eqnarray*}
A_{\mu}^a = \frac{2i}{r(r+x_0)}{\tau}_{\mu \nu}^a x_{\nu}
\end{eqnarray*}
obtained in Ref.[4]. It is obvious that each term of the
${\vec A}^a$--triplet coincides with the gauge
potential of the five--dimensional Dirac monopole with a unit topological
charge and the line of sigularity extended along the nonpositive
part of the $x_0$--axis. The vectors $A_j^a$ are orthogonal to each
other
\begin{eqnarray*}
A^a_jA^b_j = \frac{r - x_0}{r^2(r + x_0)}{\delta}_{ab}
\end{eqnarray*}
and to the vector $x_j$ ($x_jA_j^a = 0$).

By definition,
\begin{eqnarray*}
F_{ik}^a ={\partial}_iA_k^a-{\partial}_kA_i^a+
{\epsilon}_{abc}A_i^bA_k^c
\end{eqnarray*}
or, in a more explicit form,
\begin{eqnarray*}
F_{0 \mu}^a = -\frac{2i}{r^3}{\tau}_{\mu \nu}^ax_{\nu} =
-\frac{r+x_0}{r^2}A_{\mu}^a
\end{eqnarray*}
\begin{eqnarray*}
F_{\mu \nu}^a = \frac{1}{r^2}\left(x_{\nu}A_{\mu}^a -
x_{\mu}A_{\nu}^a - 2i{\tau}_{\mu \nu}^a\right)
\end{eqnarray*}
The straighforward computation gives
\begin{eqnarray}
F_{ik}^aF_{ik}^b{\hat T}_a{\hat T}_b = \frac{4}{r^4}{\hat T}^2
\end{eqnarray}
where ${\hat T}^2={\hat T}_a{\hat T}_a$.

\section {Yang SO(5) symmetry}

The YCS is governed by the Hamiltonian
\begin{eqnarray*}
{\hat H}
 = \frac{1}{2m}{\hat \pi}^2 + \frac{\hbar^2}{2mr^2}
{\hat T}^2 - \frac{e^2}{r}
\end{eqnarray*}
where ${\hat \pi}^2={\hat \pi}_j{\hat \pi}_j$,
\begin{eqnarray*}
{\hat {\pi}}_j = -i\hbar \frac{\partial}{\partial x_j} -
\hbar A_j^a{\hat T}_a
\end{eqnarray*}
and
\begin{eqnarray}
[{\hat \pi}_i,x_k] = -i \hbar {\delta}_{ik},\,\,\,\,\,
[{\hat \pi}_i,{\hat \pi}_k] = i {\hbar}^2F_{ik}^a{\hat T}_a
\end{eqnarray}
Let us consider the operator
\begin{eqnarray*}
{\hat L}_{ik} = \frac{1}{\hbar}\left(x_i{\hat \pi}_k
-x_k{\hat \pi}_i\right)-r^2F_{ik}^a{\hat T}_a
\end{eqnarray*}
It is easy to verify that
\begin{eqnarray}
[{\hat L}_{ik}, x_j] = i {\delta}_{ij}x_k - i {\delta}_{kj}x_i
\end{eqnarray}
For the commutator $[{\hat L}_{ik}, {\pi}_j]$ we have
\begin{eqnarray*}
[{\hat L}_{ik}, {\hat \pi}_j] = i {\delta}_{ij}{\hat \pi}_k -
i {\delta}_{kj}{\hat \pi}_i + {\hat Q}_{ikj}
\end{eqnarray*}
where
\begin{eqnarray*}
{\hat Q}_{ikj} = i\hbar\left(x_iF_{kj}^a - x_kF_{ij}^a\right){\hat T}_a
+ [{\hat \pi}_j, r^2F_{ik}^a{\hat T}_a]
\end{eqnarray*}
There are four possibilities for the indices $i, j, k$:
\begin{eqnarray*}
\left(\begin{array}{c}
i \\
j \\
k
\end{array} \right) =
\left(\begin{array}{cccc}
\mu & \mu & 0 & 0\\
\nu & \nu & \nu & \nu\\
\alpha & 0 & \alpha & 0\\
\end{array} \right)
\end{eqnarray*}
and, therefore, the direct calculation is required. After some algebra
we obtain ${\hat Q}_{ikj}=0$, and hence
\begin{eqnarray}
[{\hat L}_{ik}, {\hat \pi}_j] = i {\delta}_{ij}{\hat \pi}_k -
i {\delta}_{kj}{\hat \pi}_i
\end{eqnarray}
Now the commutation rule for the $SO(5)$ group generators
\begin{eqnarray}
[{\hat L}_{ij},{\hat L}_{mn}] = i {\delta}_{im}{\hat L}_{jn} -
i {\delta}_{jm}{\hat L}_{in} - i {\delta}_{in}{\hat L}_{jm} +
i {\delta}_{jn}{\hat L}_{im}
\end{eqnarray}
can be derived from (3) and (4). Moreover, it follows from (3) and
(4) that ${\hat L}_{ik}$ commutes with ${\hat H}$. This $SO(5)$
group was previously proposed by Yang [2] as the dynamical group
of symmetry for the Hamiltonian ${\hat H}_Y - e^2/r$ including only
a monopole--isospin interaction.

\section {SO(6) symmetry of YCS}

Let us consider the operator
\begin{eqnarray}
{\hat M}_k = \frac{1}{2\sqrt m}\left({\hat \pi}_i{\hat L}_{ik}+
{\hat L}_{ik}{\hat \pi}_i + \frac{2me^2}{\hbar}\frac{x_k}{r}\right)
\end{eqnarray}
by analogy with the Runge--Lenz vector [5]. Long manipulation exercises
yield $[\hat H, {\hat M}_k] = 0$, which means that ${\hat M}_k$
is the constant of motion. Now, from (3), (4) and (5) one can
show
\begin{eqnarray*}
[{\hat L}_{ij},{\hat M}_k] = i{\delta}_{ik}{\hat M}_j -
i{\delta}_{jk}{\hat M}_i
\end{eqnarray*}
More complicated calculation leads to the formula
\begin{eqnarray*}
[{\hat M}_i,{\hat M}_k] = -2i\hat H{\hat L}_{ik}-
\frac{i}{m}x_ix_kF_{mn}^a{\hat T}_a{\hat \pi}_m{\hat \pi}_n
-\frac{2{\hbar}^2}{m}\frac{x_ix_k}{r^4}{\hat T}^2
\end{eqnarray*}
It is easily to verify from (1) and (2) that last two
terms cancel each other and, therefore,
\begin{eqnarray*}
[{\hat M}_i,{\hat M}_k] = -2i\hat H{\hat L}_{ik}
\end{eqnarray*}
This commutator is identical with the corresponding commutator
for the Coulomb problem. For
${\hat M'}_i = \left(-2\hat H\right)^{-1/2}{\hat M}_i$ one has
\begin{eqnarray*}
[{\hat M'}_i,{\hat M'}_k] = i{\hat L}_{ik}
\end{eqnarray*}
Now, introduce the operators
\begin{eqnarray*}
{\hat D}_{\mu \nu} = \cases{{\hat L}_{\mu \nu}, &if $\mu,\nu=0,1,2,3,4$;
\cr
{\hat M'}_{\mu}, &if $\mu=0,1,2,3,4$; $\nu=5$; \cr
-{\hat M'}_{\nu}, &if $\mu=5$; $\nu=0,1,2,3,4$. \cr}
\end{eqnarray*}
The components of ${\hat D}_{\mu \nu}$ give an $so(6)$ algebra
\begin{eqnarray*}
[{\hat D}_{\mu \nu},{\hat D}_{\lambda \rho}] = i {\delta}_{\mu \lambda}
{\hat D}_{\nu \rho} - i {\delta}_{\nu \lambda}{\hat D}_{\mu \rho} -
i {\delta}_{\mu \rho}{\hat D}_{\nu \lambda} +
i {\delta}_{\nu \rho}{\hat D}_{\mu \lambda}
\end{eqnarray*}
Since $[{\hat H}, {\hat D}_{\mu \nu}]=0$, one concludes
that CYS is provided by the $SO(6)$ group of hidden symmetry.

\section {YCS energy spectrum}

Having obtained the group of hidden symmetry one can calculate the
energy eigenvalues by a pure algebraic method.

It is known from Ref.[6] that the Casimir operators for $SO(6)$ are
\begin{eqnarray*}
{\hat C}_2 &=& \frac{1}{2}{\hat D}_{\mu \nu}{\hat D}_{\mu \nu} \\ [2mm]
{\hat C}_3 &=& {\epsilon}_{\mu \nu \rho \sigma \tau \lambda}
{\hat D}_{\mu \nu}{\hat D}_{\rho \sigma}{\hat D}_{\tau \lambda}  \\ [2mm]
{\hat C}_4 &=& \frac{1}{2}{\hat D}_{\mu \nu}{\hat D}_{\nu \rho}
{\hat D}_{\rho \tau}{\hat D}_{\tau \mu}
\end{eqnarray*}
According to [6], the eigenvalues of these operators can be
taken as
\begin{eqnarray*}
C_2 &=& {\mu}_1({\mu}_1+4) + {\mu}_2({\mu}_2+2) + {\mu}_3^2 \\ [2mm]
C_3 &=& 48({\mu}_1+2)({\mu}_2+1){\mu}_3 \\ [2mm]
C_4 &=& {\mu}_1^2({\mu}_1+4)^2 + 6{\mu}_1({\mu}_1+4)
+ {\mu}_2^2({\mu}_2+2)^2 + {\mu}_3^4 - 2{\mu}_3^2
\end{eqnarray*}
where ${\mu}_1$, ${\mu}_2$ and ${\mu}_3$ are the positive integer
or half--integer numbers and  ${\mu}_1 \geq {\mu}_2 \geq {\mu}_3$.

The direct and very hard calculations leads to the representation
\begin{eqnarray}
{\hat C}_2 &=& - \frac{e^4m}{2{\hbar}^2\hat H} + 2{\hat T}^2 - 4 \\ [2mm]
{\hat C}_3 &=& 48\left(-\frac{me^4}{2{\hbar}^2{\hat H}}\right)^{1/2}
{\hat T}^2 \\ [2mm]
{\hat C}_4 &=& {\hat C}_2^2 + 6{\hat C}_2 - 4{\hat C}_2{\hat T}^2 -
12{\hat T}^2 + 6{\hat T}^4
\end{eqnarray}
From the last equation we can obtain another expression for the
eigenvalue $C_4$
\begin{eqnarray*}
C_4 = \left[C_2 - 2T(T+1)\right]^2 +
6\left[C_2 - 2T(T+1)\right] + 2T^2(T+1)^2
\end{eqnarray*}
and conclude that
\begin{eqnarray}
C_2 - 2T(T+1) = {\mu}_1({\mu}_1+4)     \\ [2mm]
{\mu}_2^2\left({\mu}_2 +2\right)^2 + {\mu}_3^4 - 2{\mu}_3^2 =
2T^2(T+1)^2
\end{eqnarray}
The energy levels of CYS can be obtained from (7) and (10)
\begin{eqnarray}
{\epsilon}_N^T=-\frac{me^4}{2\hbar^2({\mu}_1+2)^2}
\end{eqnarray}
The substitution of the eigenvalues of ${\hat H}$ and ${\hat T}^2$
in the equation for ${\hat C}_3$ gives one more formula for $C_3$
\begin{eqnarray*}
{C}_3 = 48({\mu}_2+2)T(T+1)
\end{eqnarray*}
Now we have two expressions for $C_3$ and the comparison leads to the
relation
\begin{eqnarray}
T(T+1) = ({\mu}_2+2){\mu}_3
\end{eqnarray}
Comparing this with (11), we have the equation
\begin{eqnarray*}
\left({\mu}_2^2 - {\mu}_3^2\right)
\left[({\mu}_2 +2)^2 - {\mu}_3^2\right] = 0
\end{eqnarray*}
Since ${\mu}_3 \leq {\mu}_2$, one concludes that ${\mu}_3={\mu}_2$.
Then, from (13) it follows that ${\mu}_2=T$, which means that ${\mu}_1$ in
(12) takes only values ${\mu}_1=T, T+1, T+2,...$.

The quantum number $T$ fixes the type of the CYS: the bosonic
type for $T=0, 1, 2,...$, and the fermionic type for
$T=1/2, 3/2,...$. The bosonic type of CYS with $T=0$ is the
five--dimensional Coulomb system.

\vspace{1cm}
{\large{\bf Acknowledgements.}}
\vspace{0.3cm}

It is a pleasure to acknowledge G. Pogosyan for helpful comments.

\end{document}